\documentclass[preprint,showpacs,preprintnumbers,amsmath,amssymb,nofootinbib]{revtex4}

\usepackage{graphicx}
\usepackage{dcolumn}
\usepackage{bm}
\usepackage{amsmath,dsfont}
\usepackage{amssymb,amsthm,amscd, amsbsy, array}
\usepackage{graphics,graphicx,xcolor}
\usepackage{etex}
\usepackage{soul} 

\numberwithin{equation}{section}

\newcommand{\bigbox}[1]{\fbox{%
\rule[-20pt]{0pt}{45pt}$\;\;\displaystyle{#1}\;\;$}
}
\newcommand{\medbox}[1]{\fbox{%
\rule[-10pt]{0pt}{25pt}$\;\;\displaystyle{#1}\;\;$}%
}

\newcommand{\half}{{\scriptstyle{\frac{1}{2}}}}
\def\2{{\half}}
\def\smallcirc{{\,\raise 0.5pt \hbox{$\scriptstyle\circ$}\,}}
\newcommand{\const}{\mathop{\rm const}\nolimits}

\def\p{{\partial}}

\newcommand{\bA}{{\bm A}}
\newcommand{\bB}{{\bm{B}}}

\newcommand{\bp}{{\bm p}}

\def\bnabla{\mbox{\boldmath$\nabla$}}

\def\bB{{\bm{B}}}

\def\beq{\begin{equation}}
\def\eeq{\end{equation}}
\def\beqa{\begin{eqnarray}}
\def\eeqa{\end{eqnarray}}
\def\nn{\nonumber}
\def\barray{\left(\begin{array}}
\def\earray{\end{array}\right)}
\def\barraynb{\begin{array}}
\def\earraynb{\end{array}}
\def\benu{\begin{enumerate}}
\def\eenu{\end{enumerate}}

\def\smallover#1/#2{\hbox{$\textstyle\frac{#1}{#2}$}} %

\def\bx{{\bm{x}}}

\def\by{{\bm{y}}}

\def\bF{{\bm{F}}}

\def\bX{{\bm{X}}}
\def\bY{{\bm{Y}}}

\newcommand{\cA}{{\mathcal{A}_{+}}}
\newcommand{\cB}{{\mathcal{A}_{\times}}}

\def\?{{\;\gb{\,?\,} \;}}

\def\bec{\begin{center}}
\def\ec{\end{center}}
\def\benu{\begin{enumerate}}
\def\eenu{\end{enumerate}}

\usepackage{color}
\newcommand{\red}{\textcolor{red}}  
\newcommand{\blue}{\textcolor{blue}}

\newcommand{\gb}{\colorbox{green}}

\newenvironment{redtext}{\color{red}}{\ignorespacesafterend} 
\newenvironment{bluetext}{\color{blue}}{\ignorespacesafterend}

\newcommand{\bblue}{\begin{bluetext}} 
\newcommand{\eblue}{\end{bluetext}} 
\newcommand{\bred}{\begin{redtext}}
\newcommand{\ered}{\end{redtext}}

\def\besub{\begin{subequations}}
\def\esub{\end{subequations}}

\begin{document}

\preprint{  
arXiv:1807.00765v6 [gr-qc] 
}

\title{Ion Traps and the Memory Effect for Periodic Gravitational Waves 
}

\author{
P.-M. Zhang${}^{1}$\footnote{e-mail:zhpm@impcas.ac.cn},
M. Cariglia${}^{2}$\footnote
{e-mail: marco.cariglia@ufop.edu.br},
C. Duval${}^{3}$\footnote{deceased},
M.~Elbistan${}^{1}$\footnote{mailto:elbistan@impcas.ac.cn.},
G. W. Gibbons${}^{4}$\footnote{mail:G.W.Gibbons@damtp.cam.ac.uk},
P.~A.~Horvathy${}^{1,5}$\footnote{
e-mail:horvathy-at-lmpt.univ-tours.fr},
}

\affiliation{${}^{1}$Institute of Modern Physics, Chinese Academy of Sciences,
Lanzhou (China)
\\
${}^{2}${\small DEFIS, Universidade Federal de Ouro Preto,  MG-Brasil},
\\
${}^3$Aix Marseille Univ, Universit\'e de Toulon, CNRS, CPT, Marseille, France
\\
${}^{4}$Department of Applied Mathematics and Theoretical  Physics,
Cambridge University, Cambridge, UK
\\
${}^{5}$ Institut Denis Poisson CNRS/UMR 7013 - Universit\'e de Tours - Universit\'e d'Orl\'eans Parc de Grandmont, 37200, Tours, (France)
}

\date{\today}

\begin{abstract}
The Eisenhart lift of a Paul Trap used to store ions in molecular physics is a linearly polarized periodic   gravitational wave. A modified version of Dehmelt's Penning Trap is  in turn related to circularly polarized periodic gravitational waves, sought for in inflationary models. Similar equations rule also the Lagrange points in Celestial Mechanics. The explanation is provided by anisotropic oscillators. 
\\[4pt]
\noindent
Phys.\ Rev.\ D {\bf 98} (2018) no.4,  044037
  doi:10.1103/PhysRevD.98.044037
\\[8pt]
\end{abstract}


\maketitle

\tableofcontents

\section{Introduction}

The \emph{Memory Effect of Gravitational Waves} concerned, originally, the motion of test particles after the passage of a sudden burst of gravitational wave. See
\cite{ZelPol,BraGri,PodSB,Favata,ShortLong,Harte,Shore,Faber,Kulczycki,Maluf,ImpMemory} and references therein for a non-exhaustive list. Later, the meaning of the expression was extended to include also the effect of periodic  gravitational waves \cite{POLPER} sought for in inflationary models \cite{Bmode,LIGOPer}.
Recent studies  \cite{POLPER,Ilderton,Andrzejewski,Tekin} reveal striking similarities with that of \emph{storing molecular ions}, considered half a century ago 
 \cite{PaulNobel,DehmeltNobel,Brown,Blumel,Blaum}.
In this paper we argue that this similarity is \emph{not a coincidence}~:
 \emph{Paul Traps} \cite{PaulNobel,Blumel} correspond indeed to \emph{Linearly Polarised Periodic (LPP) gravitational waves}; \emph{Dehmelt's Penning Trap} \cite{DehmeltNobel,Brown,Blaum} is in turn reminiscent of \emph{Circularly Polarized Periodic (CPP) gravitational waves} \cite{POLPER}, sought for in inflationary models \cite{Bmode,LIGOPer}.  
A CPP wave is also the ``double copy'' of Bia\l ynicki-Birula's electromagnetic vortex \cite{BBvort,Ilderton}.
Similar considerations apply to the Lagrange points in the 3-body problem in Celestial Mechanics \cite{BBLag,BBTroj}.

The similarity between these at first sight far remote physical phenomena, observed on so different scales, is explained mathematically by tracing back to \emph{anisotropic oscillators}. The motion of a test particle in a CPP GW
  boils down, in particular, to Hill's equations for a harmonic oscillator in a constant magnetic field.
  
Time-dependent (or not), anisotropic (or not) oscillators, described by Hill's equations and their particular case studied by Mathieu have indeed a huge literature impossible to cite here. Their general study goes beyond our scope ; here  our interest is limited to those cases which have direct relevance for the \emph{memory effect for periodic gravitational waves}. 

Apart of pointing out the far-reaching analogies 
mentioned above, we argue that applying those well-elaborated tools of ion physics to gravitational waves sheds some new light on the memory effect.
To make our paper self-contained we include some facts which are familiar for specialists of either of the fields, --- but, perhaps, not for every reader. 
 
\goodbreak
 
\section{Paul Traps}\label{Paul2d}


The intuitive explanation of the working of
Paul's ingenious ``\emph{Ionenk\"afig}''   (called now the \emph{Paul Trap}) to capture ions \cite{PaulNobel,Blumel}, has been given by Paul himself in his Nobel Lecture \cite{PaulNobel}. Let us consider indeed an electric field in the  ${X}^{+} - {X}^{-}$ plane, given by an anisotropic harmonic electric (quadrupole) potential
\beq
\Phi=\frac{\Phi_0}{2}\left(({X}^{+})^2-({X}^{-})^2\right),\qquad
{\Phi_0}=\const.
\label{saddlepot}
\eeq
 Putting $a={(e/{m})\Phi_0}$, the equations of motion of a spinless ion with charge $e$ and mass $m$ are 
\beq
\ddot{X}^{\pm} \pm a\,{X}^{\pm} = 0,
\label{xyeq}
\eeq
where the dot $\dot{(\,.\,)}$ means  $d/dt$ with $t$ denoting non-relativistic time. The opposite signs in (\ref{xyeq}) come from the relative minus sign in (\ref{saddlepot}), required by the Laplace condition $\Delta\Phi=0$ which expresses the fact that there are no sources (charges) inside the trap. For $a>0$ (say), the electric force is thus attractive in the ${X}^{+}$, and repulsive in the ${X}^{-}$ coordinate, yielding bounded oscillations in the first, but escaping motion in the second direction. 
Then Paul proposed stabilizing the position by adding a periodical perturbing electric force, i.e., to consider \footnote{The magnetic field induced by the time-varying electric field is neglected. In his Nobel lecture Paul illustrated his idea by to putting a ball on a rotating saddle surface \cite{PaulNobel}, 
materially realized in glass; a photo is reproduced in Bialynicki-Birula's lecture \cite{BBTroj}.}, 
\beq
\bF=-e\Big({\Phi_0}-\Gamma_0\cos\omega t\Big)\,
{\small \barray{r}
{X}^{+}\\
-{X}^{-}
\earray}\,,
\label{PaulForce}
\eeq
where ${\Phi_0}$ and  ${\Gamma_0}$ are constants and $\omega$ is the frequency of the  perturbation. 
The time dependent inhomogenous rf voltage changes the sign of the electric force periodically. In a certain range of parameters,  
this yields stable motions both in the $X^+$ and $X^-$ directions.


Mathematically, the planar Paul Trap is described by the modified equations
\beq
\ddot{X}^{\pm} = \mp \big(a - 2q\cos \omega t\big)\,X^{\pm}\,,
\label{XMathieu}
\eeq
where $a$ and $q=e\Gamma_0/2m$ are constants determined by the applied dc and rf voltages, respectively. 
In eqns. (\ref{XMathieu}) we recognize two uncoupled the \emph{Mathieu equations}, whose standard form is
\beq
\dfrac {d^2 \xi}{d\tau^2} + \big (a - 2 q \cos (2\tau) \big) \xi =0
\label{Matheqn}
\eeq
and whose solutions are combinations of the (even/odd) Mathieu cosine/sine functions $C(a,q,\tau)$ and $S(a,q,\tau)$, respectively.
Mathieu functions have a rather complicated behavior; in a suitable range of the parameters the solutions of (\ref{Matheqn}) remain bounded, while in another one they are unbounded. 

Returning to the eqns (\ref{XMathieu}) we note that  $\omega$,  the frequency of the oscillation, 
does \emph{not} have (as long as it does not vanish) any influence on \emph{what} will happen, only on \emph{when} will it happen.
Redefining indeed the time as 
\beq
t \to U = \half \omega t \;\;\Rightarrow\;\; \ddot{X}^{i}\to (\omega^2/4) d^2 X^{i}/dU^2\equiv (\omega^2 /4)X^{\prime\prime}
\label{tredef}
\eeq
 takes  (\ref{XMathieu})
into the standard Mathieu form (\ref{Matheqn})  with redefined parameters,
$
d^2X^{i}/dU^2\pm(\hat a - 2\hat q \cos 2U)X^{i} = 0\,,
\,
\hat a = (4/\omega^2) a,\,\hat q = (4/\omega^2 ) q.
$
Thus $\omega$  simply sets the time scale. 
 Henceforth we shall use the redefined ``time'' coordinate $U$; $d/dU$ will be denoted by prime, $(\,.\,)' =d/dU$~.  

\section{Periodic Gravitational waves}\label{PerGravWsec}

Equations similar to (\ref{XMathieu}) have been met  recently  in a rather different context,
 namely for the \emph{memory effect}, more precisely, for \emph{particle motion in the spacetime of a periodic gravitational wave} \cite{POLPER}, which is our main interest in this paper, --- 
and this is \emph{not a coincidence}, as we  now explain. 
  
A convenient way to study non-relativistic motion in $(d,1)$ dimensions with coordinates $(\bX,U)$ is indeed to consider null geodesics in  $(d+1,1)$ dimensional ``Bargmann'' space with coordinates $(\bX,U,V)$, with the potential $\Phi(\bX,U)$ entering into the $UU$ component of the metric \cite{Bargmann}. In detail, for the planar Paul Trap we have, 
\besub
\begin{align}
&d\bX^2+ 2dUdV -2\Phi(\bX,U)\, dU^2\,,
\label{Bmetric}
\\
&\Phi(\bX,U)=\half(a-2q\cos 2U)\Big({(X^{+})}^2-{(X^{-})}^2\Big)\,,
\label{PaulPot}
\end{align}
\label{Paulmetric}
\esub
whose null geodesics project to non-relativistic space-time with coordinates $(X^{\pm},U)$ precisely following eqns. (\ref{XMathieu})\,. Let us stress that the anisotropy of the profile follows from the requirement of  Ricci-flatness of the metric~: $R_{\mu\nu}=0$ for
 (\ref{Bmetric}) which implies  $\Delta \Phi=0$.
In conclusion,  the Bargmann metric of the planar Paul Trap  is \emph{an exact plane gravitational wave}. 

 More generally, an exact plane wave metric in $4D$ can be brought to the form
\besub
\begin{align}
ds^2=g_{ij} dX^i dX^j + 2dUdV + K_{ij}(U) X^i X^j dU^2\,,
\label{Brinkmetric}
\\[6pt]
K_{ij}(U){X^i}{X^j}=
\half{\cA}(U)\Big({(X^{+})}^2-{(X^{-})}^2\Big)+\cB(U)\,\Big(X^{+}X^{-}\Big)\,,
\label{Bprofile}
\end{align}
\label{genBrink}
\esub
where $\cA$ and $\cB$ are the $+$ and $\times$ polarization-state amplitudes \cite{Brinkmann,exactsol,Harte}. The geodesic equations,  
\begin{subequations}
\begin{align}
& \dfrac{d^2\bX}{dU^2} - K(U)\,\bX = 0,
\qquad\qquad
 K(U)= (K_{ij}(U)) =\half\barray{lr}
{\cA} &{\cB}\\{\cB} & -{\cA}\earray,
\label{ABXeq}
\\[10pt]
& \dfrac{d^2 V}{dU^2} + \dfrac{1}{4}\dfrac{d{\cA}}{dU\,}\left((X^{+})^2 - (X^{-})^2 \right )
+{\cA}\left(X^{+}\dfrac{dX^{+}}{dU\,} - X^{-}\dfrac{dX^{-}}{dU\,} \right )
\nn
\\[6pt]
&\qquad+\dfrac{1}{2} \dfrac {d{\cB}}{dU\,} X^{+}X^{-}
+ {\cB}\left(X^{-}\dfrac{dX^{+}}{dU\,} + 
X^{+}\dfrac{dX^{-}}{dU\,} \right)
= 0\,,
\label{ABVeq}
\end{align}
\label{ABeqs}
\end{subequations}
are decoupled~: after solving (\ref{ABXeq}) for the transverse motion, (\ref{ABVeq}) can be integrated. 

Eqns. (\ref{ABXeq}) belong to family of Hill-type equations which describe (possibly time-dependent and/or anisotropic) oscillators. Their general study goes well above our scope here. 
Having established the fundamental relation we focus henceforth our study to those cases which are directly relevant for us -- namely to the motion of test particles initially at rest in a circularly polarized gravitational wave. 

Henceforth we focus our attention at the transverse motion.

\begin{itemize}
\item
The Bargmann metric of the Paul Trap,  (\ref{Paulmetric}), is a \emph{linearly polarized gravitational wave with periodic profile}. Its  properties for $a=0$, i.e. for the periodic profile
\beq 
\cA =A_0\cos2U,\,\quad \cB =0
\label{linpolperiodic}
\eeq 
were studied in \cite{POLPER} 
(see also e.g. \cite{Harte}), shown in Fig.\ref{linpolper} below. 
For particular values of the parameters, one obtains bound motions. The intuitive explanation is precisely that of Paul recalled in sec. \ref{Paul2d}~: in a given ``moment'' $U$ one of the oscillators is attractive and the other is repulsive, with strength $\cA=A_0\cos2U$. However as ``time'' goes on, the strength varies, and when the cosine changes sign, the attractive and repulsive sectors are interchanged, as we told in sec. \ref{Paul2d}.
\begin{figure}[h]
\begin{center}
\includegraphics[scale=.24]{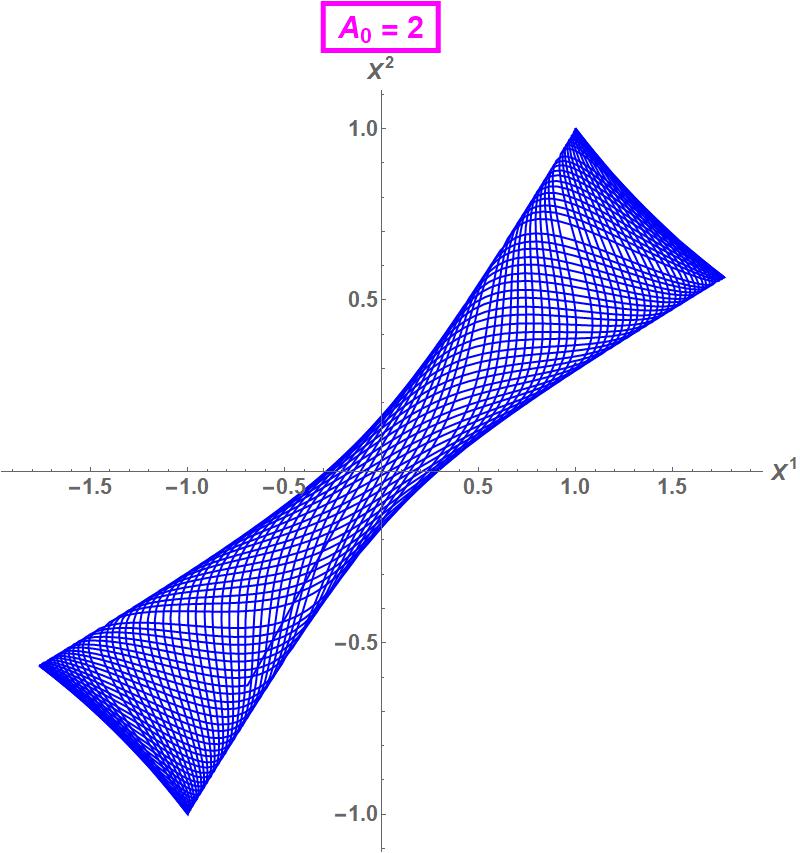}
\\
\end{center}\vskip-8mm
\caption{\textit{\small 
In a weak linearly polarized periodic (LPP) wave, (\ref{linpolperiodic}), the transverse coordinate $\bX(U)$ 
 oscillates in a bounded ``bow tie''-shaped domain.
 The initial conditions are $\dot X^+ (U\!=\!0) = \dot X^- (U\!=\!0) = 0$ (at rest for $U\!=\!0$), at initial position $X^+(U\!=\!0)= 1,\, X^-(U\!=\!0) = 0$.
\label{linpolper}}}
\end{figure}

A sufficiently strong wave breaks up the bound motion.

\goodbreak
\item
The general form in eqn. (\ref{Bprofile}) allows however also for more general profiles, and now we turn to waves with  \emph{circularly polarized periodic profile} (CPP), considered before e.g. in \cite{POLPER},  \goodbreak
\beq
K = \big(K_{ij}) = \frac{A_0}{2} \barray{lr} 
\cos 2U & \sin 2U
\\
\sin 2U &- \cos 2U
\earray
\qquad
A_0=\const>0\,.
\label{circperprof}
\eeq
\end{itemize}

\begin{figure}[ht]
\begin{center}
\includegraphics[scale=.2]{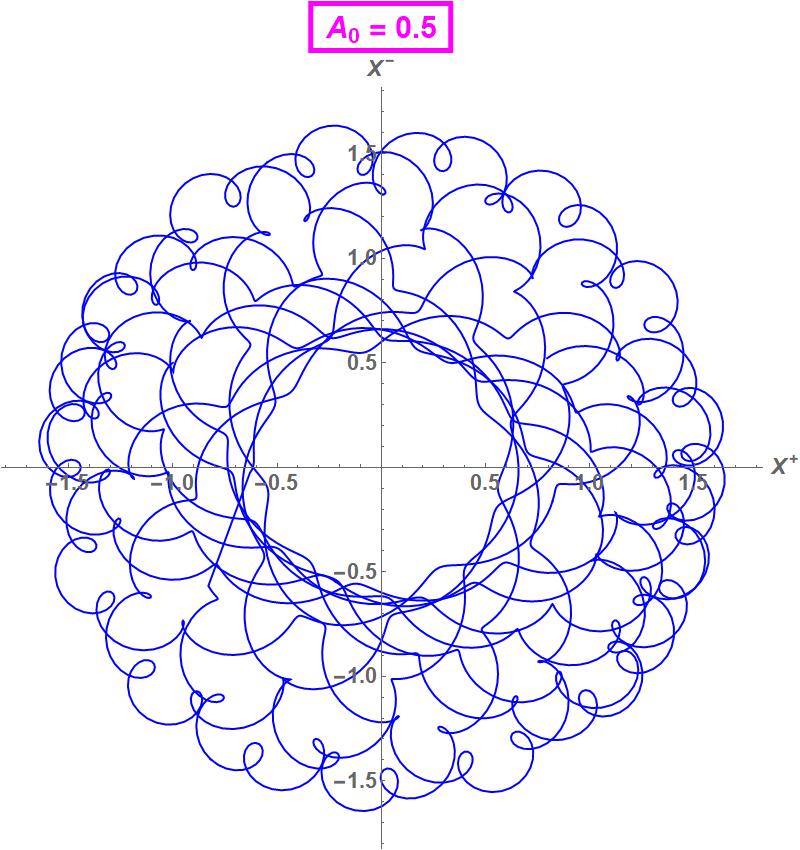}
\qquad\qquad
\includegraphics[scale=.19]{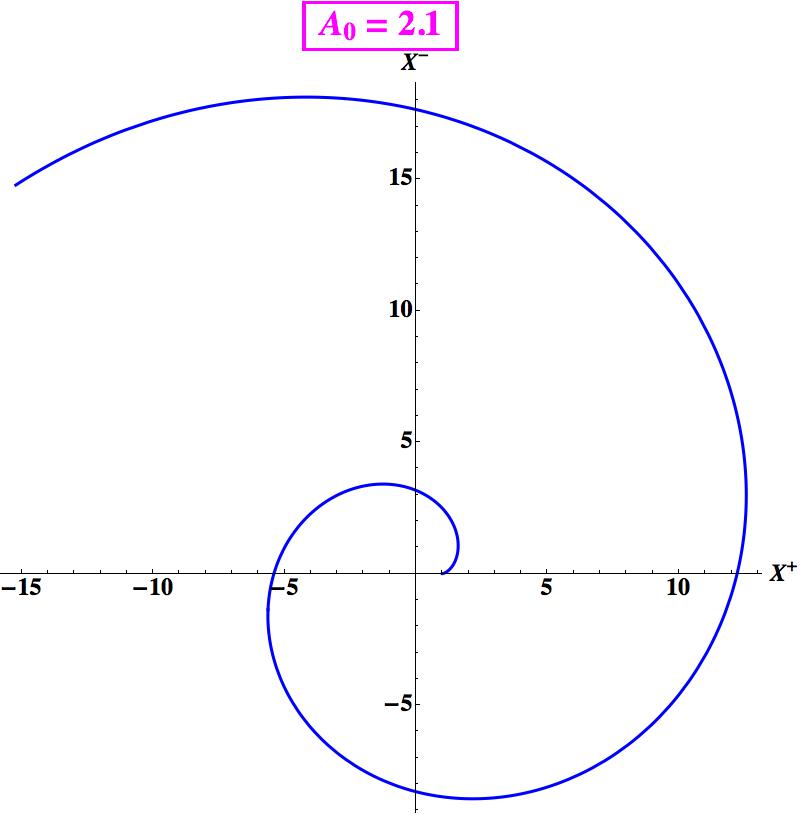}
\vskip-8mm
\null\hskip-1mm{(i)}\hskip70mm (ii)
\\
\vskip-2mm
\caption{\textit{\small (i) In a sufficiently weak circularly polarized gravitational wave (\ref{circperprof}) the transverse 
 trajectory 
  of a particle initially at rest remains confined in a toroidal region. 
(ii) For a strong wave the trajectory becomes unbounded.
The  initial conditions are  
$\dot X^+ (U=0) = \dot X^- (U=0) = 0$
and
$X^+ (U=0) = 1,\,  X^-(U=0) = 0.$
} 
\label{cirper}}
\end{center}
\end{figure}

The transverse eqns of motion,
$ \bX^{\prime\prime}=K\bX,
$ should be supplemented by appropriate initial conditions. In the sandwich case one usually considers particles which are at rest in the before zone. But  a periodic wave  has no before zone, and  here we propose the  initial condition  
\footnote{Ions issued from accelerators and injected into the ``Ionenk\"afig"  require different initial conditions.}, 
\beq 
\text{rest\  at\ $U=0$} \quad\text{i.e.,}\quad
\bX^{\prime}(0)=0.
\label{restcond}
\eeq 
 Then numerical calculations  \cite{POLPER}  
 yield Fig.\ref{cirper}~: for a sufficiently weak wave all motions remain confined to a toroidal region; for a strong wave  the trajectory becomes instead unbounded~: the particle is ejected.
Below we show that the problem admits  an \emph{exact analytic solution}. 
Following a suggestion of Kosinski \cite{Kosinski}, the first step is to switch to a rotating frame by setting
\beq
\barray{c}X^+\\ X^-\earray
=
\barray{rcr}\cos U &
&-\sin  U\\
\sin U & &\cos U\earray
\barray{c}Y^+\\ Y^-\earray.
\label{rotframe}
\eeq
In terms of the new coordinates $Y^{\pm}$ the  harmonic force becomes $U$-\emph{independent} --- at the price of introducing the cross terms $\mp2(Y^{\mp})^{\prime}$  \footnote{In $Y$-coordinates $U$-translational symmetry is  restored due to the \emph{manifest $U$-independence of the metric (\ref{Ymetric})}. Expressed in  the original coordinates, the 6th ``screw'' symmetry  \cite{exactsol,Sippel,Carroll4GW,POLPER,
Ilderton} is recovered.}, 
\beq\medbox{
(Y^{\pm})^{\prime\prime} \mp 2(Y^{\mp})^{\prime}-  {\Omega^2_{\pm}}\, Y^{\pm} = 0\,
\quad\text{where}\quad
\Omega_{\pm}^2={1\pm A_0/2}\,.}
\label{Yeqn}
\eeq

Our  initial condition (\ref{restcond}) is valid in Brinkmann-coordinates (\ref{Brinkmetric}) ; from Eqn. (\ref{rotframe}) we infer instead
\beq
\bY^{\prime}(0) =
\barray{cc} 0 &1\\ -1 &0\earray
\bY_0\,= 
\barray{cc} 0 &1\\ -1 &0\earray
\bX_0\,,
\label{Y0}
\eeq
i.e., $\bY^{\prime}(0)$ is obtained  from $\bY(0)=\bX_0\,$ by a $90$-degree rotation, which corresponds precisely to rotating the coordinate system.

\goodbreak


The eqns of motion can be conveniently solved by \emph{chiral decomposition} \cite{Alvarez, ZGH}. Eqns. (\ref{Yeqn}) belong indeed to a Hamiltonian system in the plane, whose phase space is thus $4$ dimensional ; it has coordinates $Y^{\pm}$ and $\Pi^{\pm}={(Y^{\pm})}^{\prime}$. Then the idea is to choose ``smart'' phase-space coordinates we denote here by $Z_{+}^{a},\, Z_-^{b},\, a, b = 1,2 $ such that the system decouples onto uncoupled $1D$ oscillators \cite{Alvarez, ZGH}. Searching for real coefficients $\alpha_\pm$ and 
$\beta_\pm$,
\besub
\begin{align}
\label{polperdd1}
\Pi^+&=\alpha_+ Z_+^2+\alpha_- Z_-^2\,,  
&\Pi^-&=-\beta_+Z_+^1-\beta_-Z_-^1\,, 
\\
Y^+&= Z_+^1+Z_-^1\,, 
&Y^-&= Z_+^2+Z_-^2\,,
\label{polperdd2} 
\end{align}
\label{polperdd}
\esub
in terms of which both the symplectic form and the Hamiltonian separate, we find that for
$ 
\alpha_+=1, \, \alpha_-= \Omega_-^2, \, \beta_+= \Omega_+^2, \, \beta_-=1,
$ 
for example \footnote{Both the symplectic structure and the Hamiltonian are proportional to the wave amplitude, $A_0$, which drops seemingly out therefore from the equations of motion. It
is however still  hidden in the frequencies 
$\Omega_{\pm}$, cf. (\ref{Yeqn}).},
\besub
\begin{align}
\label{polpersymp}
\sigma&=\sigma_+-\sigma_-=-\frac{A_0}{2}\Big[dZ_+^1\wedge dZ_+^2-dZ_-^1\wedge dZ_-^2\Big],
\\[10pt] 
H&=H_+-H_-=\frac{A_0}{4}\Big[\big(\Omega_+^2Z_+^1 Z_+^1+Z_+^2 Z_+^2\big)-\big(Z_-^1 Z_-^1+\Omega_-^2Z_-^2 Z_-^2\big) \Big].
\label{polperH}
\end{align}
\label{polpersH}
\esub
The relative negative signs between the terms reflect here the chiral nature~: the two oscillators turn in the opposite direction \cite{ZHAGK}.
The  Poisson brackets associated with the symplectic structure (\ref{polpersymp}) are,
\beq
\{Z_+^1, Z_+^2 \}=\frac{2}{A_0}, \qquad \{Z_-^1, Z_-^2 \}=-\frac{2}{A_0}, 
\qquad
\{Z_+^1, Z_-^2 \}=\{Z_+^2, Z_-^1\}=0.
\label{polperPB}
\eeq
Working out the Hamilton equations, we end up with uncoupled oscillator equations,
\beq
{(Z_{\pm}^a)}^{\prime\prime}+\Omega_{\pm}^2 Z_+^a=0, 
\label{decompoeqn} 
\eeq 
($a=1,2$), 
whose solutions (when none of the $\Omega_{\pm}$ vanishes  \footnote{The choice of the coefficients is not unique; another choice would interchange $\Omega_+$ and $\Omega_-$. When one of the $\Omega_\pm$s vanishes the corresponding motion is free \cite{ZGH,ZHAGK,POLPER}.}), are,
\besub
\begin{align}
Z_+^1&= A\cos(\Omega_+U)+B\sin(\Omega_+ U), 
\\
Z_+^2&=-\Omega_+\Big( A\sin(\Omega_+ U) - B\cos(\Omega_+ t)\Big), 
\\
Z_-^1&= C\cos(\Omega_- U)+D\sin(\Omega_- U), \\
Z_-^2&= -\frac{1}{\Omega_-}\Big(C\sin(\Omega_-U)- D\cos(\Omega_- U)\Big), 
\end{align}
\label{chirsolutions}
\esub
where $A, B, C, D$ are constants. 
Proceeding backwards we obtain, using (\ref{polperdd2}), 
\besub
\begin{align}
Y^+(U)&=
\qquad \;\, A \cos \Omega_+ U + B \sin \Omega_+ U \;\;\;+ \qquad C \cos \Omega_- U + D \sin \Omega_- U\,,
\\[4pt]
Y^-(U)&=
  - \Omega_+ (A\sin\Omega_+ U - B\cos\Omega_+ U) \; -\;
\dfrac{1}{\Omega_-}(C\sin\Omega_- U - D \cos \Omega_- U)\,.
\end{align}
\label{polperLsolns}
\esub
For a weak wave i.e. such whose amplitude is $A_0<2$ both frequencies $\Omega_{\pm}$ in (\ref{Yeqn}) are real, implying that the motion, although complicated, remains bounded. A typical trajectory is shown in Figs.7 and 8 of \cite{POLPER}.
However for a strong wave with amplitude $A_0>2$ one (and only one) of the $\Omega_{\pm}$ becomes imaginary and the corresponding motion is unbounded: a sufficiently strong wave ejects the particle and makes it escape.
Between those two regimes i.e., for $A_0=2$, one of the $\Omega$'s vanishes, and the motion in the corresponding direction is free; we recover eqn. \# (5.11) of \cite{POLPER}, illustrated in Fig.9 of that paper. 

\begin{figure}[ht]
\begin{center}
\includegraphics[scale=.18]{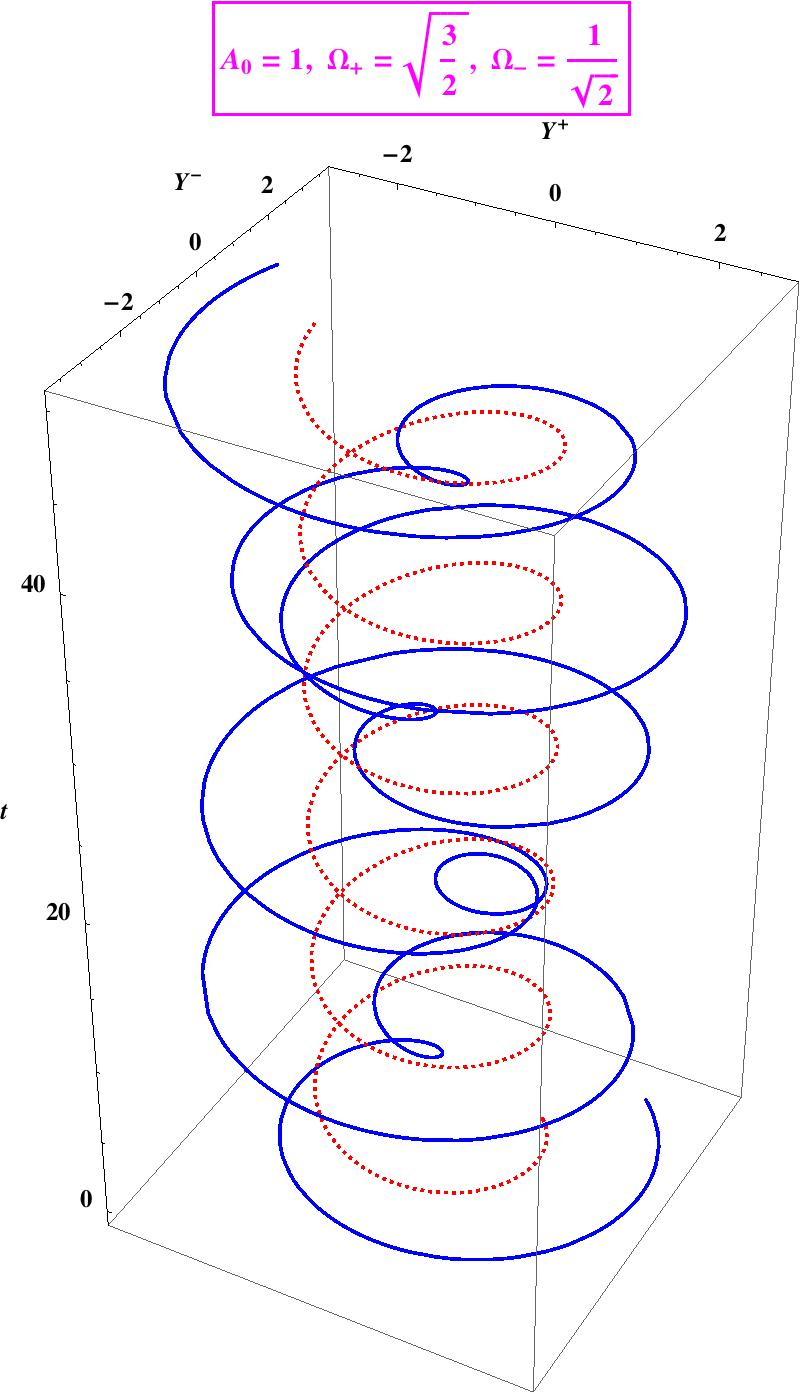}
\vskip-4mm
\caption{\textit{\small In the circularly polarized periodic gravitational wave (\ref{circperprof}) the trajectory unfolded into ``time''(\blue{\bf {in heavy blue}}) winds about the guiding center (\red{\bf dotted in red}). If the wave is weak, $A_0<2$,
 the trajectory remains bounded, projecting to the plane consistently with Fig.\ref{cirper}. 
}
\label{chiralper}} 
\end{center}
\end{figure}

The solutions (\ref{polperLsolns}) are plotted  
for $A_0 < 2$ in Fig.\ref{chiralper}.  
The one which has smaller real (or imaginary) frequency can be viewed (somewhat arbitrarily) as a {guiding center}, around which the one with the larger frequency winds around. 
For $A_0=2$ one of the frequencies vanishes, $\Omega_-=0$, and the $Y$-trajectory is an ellipse drifting with constant speed \cite{ZGH,ZHAGK}.
A further rotation (\ref{rotframe}) backward would yield the trajectories $X^{\pm}(U)$.

As noticed by Ilderton \cite{Ilderton}, eqns (\ref{Yeqn}) are actually identical to eqns. \# (10a-b) of Bia\l ynicki-Birula for a charged particle in the field of an electromagnetic vortex \cite{BBvort}, and   (\ref{polperLsolns}) above just reproduces his solution \# (14) -- with some additional insight, though. The relation will be further discussed elsewhere.   


So far we studied classical motions only. However the system could  readily be quantized, courtesy of the chiral decomposition \cite{Alvarez,ZGH,ZHAGK,BBvort}.
 The Poisson brackets  (\ref{polperPB}) are promoted to commutation relations,
\beq
\label{polpercc1}
[Z_+^i, Z_+^j ]=\frac{2i\hbar}{A_0}\epsilon^{ij}, 
\qquad [Z_-^i, Z_-^j]=-\frac{2i\hbar}{A_0}\epsilon^{ij},
\eeq
where we denoted, with a light abuse of notations, the classical and quantum observables by the same symbols. Creation and annihilation operators can now be introduced,
\besub
\begin{align}
\label{polpercladder1}
a^\dagger&=\sqrt{\frac{A_0\Omega_+}{4}}\left(Z_+^1-\frac{i}{\Omega_+}Z_+^2\right), \qquad a=\sqrt{\frac{A_0\Omega_+}{4}}\left(Z_+^1+\frac{i}{\Omega_+}Z_+^2\right)\,,
\\[8pt]
\label{polpercladder2}
b^\dagger&=\sqrt{\frac{A_0\Omega_-}{4}}\left(Z_-^2-\frac{i}{\Omega_-}Z_-^1\right), \qquad b=\sqrt{\frac{A_0\Omega_-}{4}}\left(Z_-^2+\frac{i}{\Omega_-}Z_-^1\right)\,,
\end{align}
\label{polpercladder}
\esub
whose non-vanishing commutators are, by (\ref{polpercc1}),
\beq
[a, a^\dagger]=1=[b, b^\dagger].
\eeq
 In their terms the Hamiltonian is,
\beq
\label{ploperspec}
H=\left(\Omega_+\big(a^\dagger a+\frac{1}{2}\big)-\Omega_-\big(b^\dagger b+\frac{1}{2}\big)\right).
\eeq
The number operators
$a^\dagger a$ and $b^\dagger b$ commute and have [$\hbar$-times] integer eigenvalues. The bound-state spectrum is therefore,
\beq
E_{n_+,n_-}=\hbar\Big[\Omega_+ \big(n_++\half\big) 
- \Omega_- \big(n_-+\half\big)\Big]
\qquad
n_\pm=0, 1\dots
\label{perspectrum}
\eeq
Let us observe that the spectrum is not bounded from below, consistently with the relative minus sign of $H_+$ and $H_-$ in the Hamiltonian (\ref{polperH}) reflecting the shape of the saddle potential.

\section{Sturm-Liouville problem 
\& switching to BJR}\label{SLsec}
 
The key to study the memory effect for gravitational waves is to solve the Sturm-Liouville equation with an auxiliary condition \cite{ShortLong,Torre,SLC},
\besub
\begin{align}
{P}^{\prime\prime}(U)&=K(U)P(U),
\label{SL}
\\
P(U)^T{P}^{\prime}(U)&=\big({P}^{\prime}(U)\big)^TP(U).
\label{SLaux}
\end{align}
\label{SLfull}
\esub
%
This system should be supplemented by initial conditions. Let us recall that in the sandwich case,
for which the wave vanishes outside an interval $[U_i,U_f]$, we required in the before zone $U\leq U_i$ the initial conditions
\beq
P(U)=\mathbb{I},\quad\&\quad
P^{\prime}(U)=0 
\quad\text{for}\quad U\leq U_i\,.
\label{Pinitcond}
\eeq
Below we extend our study to the periodic case, which has no before zone. 
First we note that having solved the SL eqn. (\ref{SLfull}) for the $2\times2$ matrix $P(U)$, 

\begin{enumerate}

\item
%
allows us  to switch to \textit{Baldwin-Jeffery-Rosen (BJR) coordinates} $(\bx,u,v)$~: setting
\besub
\begin{align} 
X^i &= P^{ij}(u)\, x^j \, ,
\label{Xx} 
\\ 
U &= u \,
\label{Uu}  
\\ 
V &= v - \frac{1}{4} \,y^i {(G^{ij})}^{\prime}(u)\, x^j \, , \quad
\text{where}\quad G = P^T P \, 
\label{Vv} 
\end{align} 
\label{BBJR}
\esub
carries the metric (\ref{Brinkmetric}) with $g_{ij}=\delta_{ij}$ to the BJR form
\beq
G_{ij}(u)dx^idx^j+2dudv\,.
\label{BJRmetric}
\eeq

\item
The metric admits a $5$ parameter isometry   \cite{BoPiRo,Sou73,exactsol,Sippel,Torre,Carroll4GW}. 
The system is in particular symmetric with respect to
translations and boosts, with associated conserved momenta 
\besub
\begin{align}
p_i &= G_{ij}\dot{x}^j
\label{consBJRp}
\\
k_i &=x_i(u)-H_{ij}(u)\,p_j
\label{consBJRk}
\end{align}
\esub
where $H(u)$ is the $2\times2$ matrix $H(u)=\displaystyle\int_{u_0}^u{\!G^{-1}(w)dw}$ \cite{Sou73,Carroll4GW,ShortLong}\,. 

\item
Remember that in the sandwich case the usual assumption is that the particle is at rest in the before zone, $\bX'(U)=0$ for $U<U_i$. Then 
 exporting  to BJR by (\ref{Xx}),
$$
\bx=P^{-1}\bX \quad \Rightarrow\quad \bx'(0)= \big(-P^{-1}P'(P^{-1}\big)(0)\bX(0)+(P^{-1})\bX'(0)
=0\,,
$$
and thus the BJR coordinate also has vanishing initial velocity, $\bx'(u)=0$. Consequently  the linear momentum vanishes, $\bp=0$ by (\ref{consBJRp}) ; then 
(\ref{consBJRk}) implies that $\bx(u)=\bx_0$ for all $u$. Returning to Brinkmann coordinates allows us to conclude, using (\ref{Pinitcond}), that the trajectory is simply
\beq\medbox{
\bX(U) = P(U)\bX_0\,. 
}
\label{Btraj}
\eeq
Now we extend our theory  by  replacing the initial conditions (\ref{Pinitcond}) by  requiring that it holds at a chosen initial moment, e.g.,
\beq
\medbox{
P(0)=\mathbb{I},\quad\&\quad
P^{\prime}(0)
=0 \,.}
\label{PerPinit}
\eeq 
Then  (\ref{Btraj})
remains true also in our case~: the SL eqns (\ref{SLfull}) imply that it satisfies the
equations of motion with the initial conditions $\bX(0)=\bX_0$ and $\bX'(0)=0$. Conversely, following the same argument as in the sandwich case, we observe that inverting (\ref{Xx}) shows that $\bX'(0)=0$ implies $\bx'(0)=0$ and therefore $\bp=0$ by (\ref{consBJRp}) from which (\ref{consBJRk}) allows us to infer $\bx(u)=\bx_0=\const$, so that (\ref{Xx}) yields once again (\ref{Btraj}).

\end{enumerate}

\subsection{Linearly Polarized Periodic (LPP) waves}
\label{LPPSL}

In the linearly polarized case (\ref{linpolperiodic}) 
{\tt Mathematica}  tells us that
eqn (\ref{SL}) can be solved~: Using the shorthands  $c_{A_0}(U)\equiv  C({0,A_0},U)$ and $s_{A_0}(U)\equiv S({0,A_0},U)$\, cf. sec. 2, we get, 
\begin{equation}
P(U)=
\barray{ccc}
A_{11}\,c_{A_0}(U)+B_{11}\,s_{A_0}(U)
&&A_{12}\,c_{A_0}(U)+B_{12}\,s_{A_0}(U)
\\[6pt]
A_{21}\,c_{-A_0}(U)+B_{21}\,s_{-A_0}(U)
&&A_{22}\,c_{-A_0}(U)+B_{22}\,s_{-A_0}(U)
\earray 
\label{Psol}
\end{equation}
with $A_{ij}$ and $B_{ij}$ constants of integration. 
Then 
eqn (\ref{SLaux}) yields the compatibility constraints
\begin{equation}
A_{11}B_{12}=A_{12}B_{11}
\qquad
\&
\qquad
A_{22}B_{21}=A_{21}B_{22}.
\label{Constraints}
\end{equation}
Assuming that, e.g., $A_{11}\neq0$ and $A_{22}\neq0$ we obtain $B_{12}$ and $B_{21}$. The solution  thus depends on $6$ integration constants. 
Then it follows that 
 from the parity-properties of the Mathieu functions that the initial condition 
$\bX^{\prime}(U)=0$ in (\ref{restcond}) can only be satisfied if all $B_{ij}$ vanish (and then the auxiliary conditions (\ref{Constraints}) hold also). Then, consistently with eqn (IV.3) of \cite{POLPER},
 the trajectory is  given by pure Mathieu cosines with labels $\pm A_0$  and coefficients  depending on the initial conditions,
\beq
X^{\pm} (U) = D^{\pm}\, c_{{\pm}A_0}(U)\,,
\label{PartMathSol}
\eeq
where the constants $D^{\pm}$ are determined by the  $A_{ij}$ in (\ref{Psol}) and the initial position $\bX_0$.

\subsection{Circularly Polarized Periodic (CPP) waves}\label{CPPSL}

Now we turn to the circularly polarized wave (\ref{circperprof}). Switching to a rotating frame by (\ref{rotframe}) allows us to present the metric  as, 
\besub
\begin{align}
ds^2&=
d\bY^2
+2dU\Big(dV+A\Big)
\; - \; 2\Psi dU^2,
\label{gengyrmet}
\\
A&=-Y^-dY^+ + Y^+dY^-\,,
\qquad
\Psi=- \half\Big(\Omega_+^2(Y^+)^2+\Omega_-^2(Y^-)^2\Big).
\end{align}
\label{Ymetric}
\esub
The only non-vanishing component of the Ricci tensor of (\ref{ppmetric}) is
\beq
R_{UU}=-\p_U(\bnabla\cdot\bA)-\half B^2-\Delta\Psi
\label{ppRicci}
\eeq
where $B=\p_iA_j-\p_jA_i$. Ricci-flatness is thus confirmed for (\ref{Ymetric}) \footnote{This is hardly surprising~: switching from $X$ to $Y$ is a mere coordinate change.}. 

This metric is consistent with the Bargmann description of a particle with charge = mass in a combined anisotropic oscillator plus a ``magnetic'' (alias Coriolis) field. 
The appearance of the new metric component implies that the potential term $\Psi$ alone does not contain  all information. 
%
The metric (\ref{Ymetric})  has the form of a pp metric sometimes called ``gyratonic'' \footnote{Considered by Brinkmann back in 1925 \cite{Brinkmann} and used
e.g. in \cite{DHP2}. Here we deliberately changed our notations, $\Phi \to \Psi,\, K\to H,$ to underline the difference with the previous discussion.}, 
\beq
d\bY^2+ 2dU\big(dV+A\big)-2\Psi\, dU^2,
\label{ppmetric}
\eeq
where now $\Psi= -\half H_{ij}Y^iY^j$ and
where the $1$-form $A=A_\mu dY^\mu$ is a vector potential. It has  a gauge freedom:
 $A_i \rightarrow A_i - \p_i \Lambda$ can be compensated by the ``vertical'' coordinate transformation $V \rightarrow V + \Lambda(\bY)$. 

Switching to BJR coordinates by replacing $\bX$ by $\bY$ and $\bx$ by $\by$ in (\ref{Xx}), the new term in (\ref{ppmetric}) becomes \vspace{-4mm}
\beqa
&& 2 dU A_j(Y) dY^j = 2 du \, d \left( A_j(Y) Y^j \right) - 2 du \, \p_i A_j (Y) dY^i Y^j =\nn
 \\ 
&&2 du \, d\left(A_j(Y) Y^j\right) - 2 \left({(P^{\prime})}^T \p A P \right)^{(ij)}\!\! y^i y^j du^2 - 2 \big(P^T \p A P \big)^{ij}y^j du\, dy^i \, ,
\nn 
\eeqa
where we used the shorthand $\p A$ for the matrix  $[\p A_j/\p Y_i]$. 
The first term here can be reabsorbed into the $V$-change in (\ref{Vv}),
\beq 
V = v - \frac{1}{4} y^i {(a^{ij})}^{\prime}(u) y^j - 2 A_j(Y) P^{jk} y^k  \, .
\nn 
\eeq 
The two other terms modify the Sturm-Liouville equations (\ref{SLfull})\footnote{Our formulas are valid in $4D$.}~: the auxiliary condition (\ref{SLaux})  becomes
\beq 
P^T {P^{\prime}} - {(P^{\prime})}^T P = 2 \left( P^T \p A P \right) \, , 
\nn
\eeq
whose consistency requires $\p A = - \p A^T$. 
 Then the SL equation  (\ref{SL}) becomes  
\beq 
- \frac{1}{2} \left( {\big(P^{\prime\prime}\big)}^T P + P^T P^{\prime\prime} \right) + P^T H P = {(P^{\prime})}^T \p A P - P^T \p A {P^{\prime}}.  
\nn
\eeq
When $\p_U(\p A) = 0$, both equations are solved by 
$ 
{P}^{\prime\prime} = K P + 2 \p A {P^{\prime}} \, . 
$
To sum up, changing our notations to emphasise that the new system concerns the metric obtained after applying the rotational trick, 
$\bX \to \bY\,,\, K=\big(K_{ij}\big)  \to
H=\big(H_{ij}\big)$ and $P \rightarrow Q$,
eqns (\ref{BBJR}) for the Brinkmann $\Leftrightarrow$ BJR transcription should be replaced by 
\besub
\begin{align} 
Y^i &= Q^{ij}(u) y^j \, ,
\label{AYy} 
\\ 
U &= u \,
\label{AUu}  
\\ 
V &= v - \frac{1}{4} y^i {(G^{ij})}^{\prime} y^j - 2 A_j\, Q^{jk} y^k  \,, \qquad
G = Q^T Q \, .
\label{AVv} 
\end{align} 
\label{ABBJR}
\esub
Note that the eqns (\ref{Xx}) are formally unchanged while (\ref{Vv}) picks up a new term, however the SL eqns to be solved  are now rather 
\vspace{-3mm}
\besub
\begin{align} 
Q^{\prime\prime} &=H Q+ 2 \p A \,{Q^{\prime}}  
\label{newSL} \, , 
\\ 
\p A &= - \p A^T \, ,
\label{newconstraint} 
\\ 
\p_U( \p A) &= 0 \, .
\label{whatisthis}
\end{align} 
\label{newSLfull}
\esub 
\vskip-10mm 
Spelling out our formulae for  the
circularly polarized periodic wave, from (\ref{Ymetric}) we infer that 
\beq
\p A= \barray{cc}0 &1\\ -1& 0\earray,
\qquad
H=\barray{cc}
\Omega_+^2 &0\\ 0& \Omega_-^2 
\earray\,,
\label{ourpA}
\eeq
which are both $U$-independent. Thus
our modified Sturm-Liouville equation becomes
\beq
\bigbox{
Q^{\prime\prime} + 2\barray{cc}0 &-1\\ 1& 0\earray Q^{\prime}- \barray{cc}
\Omega_+^2 &0\\ 0& \Omega_-^2 
\earray\, Q = 0\,
\\
}
\label{ournewSL}
\eeq
which is precisely eqn. (\ref{Yeqn}) with the {vector} $\bY$ replaced by the $2\times2$ 
{matrix} $Q$. Replacing $\bX$ by $\bY$ in (\ref{Btraj}), it follows that
\beq\medbox{
\bY(U)=Q(U)\bY_0
}
\label{ABtraj}
\eeq
is a solution of the equations of motion (\ref{Yeqn}). Moreover, the initial condition
(\ref{Y0}) is satisfied provided\footnote{$Q^{\prime}(0)$ is the matrix of a planar rotation by $\pi/2$.},
\beq\medbox{
Q(0)=\mathbb{I}\,,
\qquad
Q^{\prime}(0)=\barray{cc} 0 &1\\ -1 &0\earray\,.
}
\label{AP0}
\eeq

Our new equation (\ref{ournewSL}) has constant coefficients and can be solved analytically.  
 Putting $Q=(Q_{ij})$   (\ref{ournewSL})
 is mapped indeed into two sets of equations of type \eqref{Yeqn}, with the identifications $Q_{11} \leftrightarrow Y^+$, $Q_{21} \leftrightarrow Y^-$, $Q_{12} \leftrightarrow Y^+$, $Q_{22} \leftrightarrow Y^-$~: the  columns of $Q$ are vectors of the form {\tiny $\left( \begin{array}{c} Y^+ \\ Y^- \end{array} \right)$}  both of which satisfy \eqref{Yeqn}. Therefore  the general solution is, by \eqref{polperLsolns}, a combination with eight constants $A_i, \dots , D_i$, $i=1,2$ cf. (\ref{Psol}),  
\begin{subequations}
\begin{align}
Q_{11}(U)&=
\, A_1 \cos \Omega_+ U + B_1 \sin \Omega_+ U + C_1 \cos \Omega_- U + D_1 \sin \Omega_- U\,,  
\\[4pt]
Q_{21}(U)&=
  - \Omega_+ (A_1\sin\Omega_+ U - B_1\cos\Omega_+ U) \; -\;
\dfrac{1}{\Omega_-}(C_1\sin\Omega_- U - D_1 \cos \Omega_- U)\,,  
\\[4pt] 
Q_{12}(U)&=
\, A_2 \cos \Omega_+ U + B_2 \sin \Omega_+ U + C_2 \cos \Omega_- U + D_2 \sin \Omega_- U\,,  
\\[4pt]
Q_{22}(U)&=
  - \Omega_+ (A_2\sin\Omega_+ U - B_2\cos\Omega_+ U) \; -\;
\dfrac{1}{\Omega_-}(C_2\sin\Omega_- U - D_2 \cos \Omega_- U)\,,  
\end{align}
\label{genQsol}
\end{subequations} 
The number of constants is halved by the initial condition \eqref{AP0} which require,  
\beq
C_1 = 1 - \Omega_+^2 A_1 \, , \;\;
D_1 = - \frac{\Omega_+}{\Omega_-} B_1 \, , \;\;
C_2 = - \Omega_+^2 A_2 \, , \;\; 
D_2 = \frac{1}{\Omega_-} - \frac{\Omega_+}{\Omega_-} B_2\,.
\nn
\eeq 
Requiring in addition also $Q(0) = \mathbb{I}$ 
which follows from \eqref{ABtraj} eliminates all constants with the exception of $B_2 = \Omega_+^{-1}$, leaving us with \footnote{The transverse metric in BJR form, $G_{ij}(u) = Q^T(u)Q(u)$ is not illuminating and is therefore omitted.}   
\beq 
Q = \left(\begin{array}{ccc} \cos (\Omega_- U) &\;& \frac{\displaystyle{\sin(\Omega_+ U)}}{\Omega_+} 
\\[10pt]
 - \frac{\displaystyle{\sin(\Omega_- U)}}{\Omega_-} &\;& \cos(\Omega_+ U)  \end{array} \right) \, . 
\eeq 

\section{Ion Traps in 3D}

\subsection{Paul Trap in 3D}

\begin{figure}
\includegraphics[scale=.28]{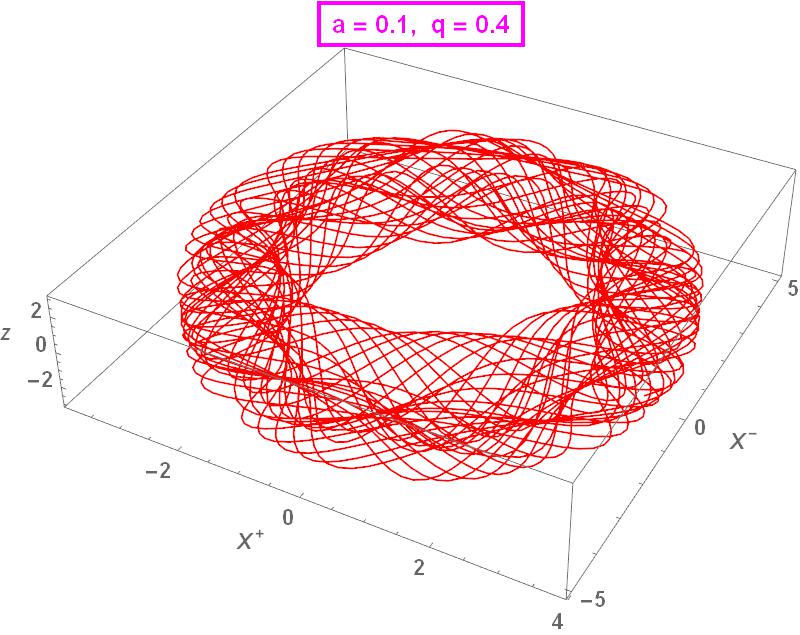}\;
\includegraphics[scale=.28]{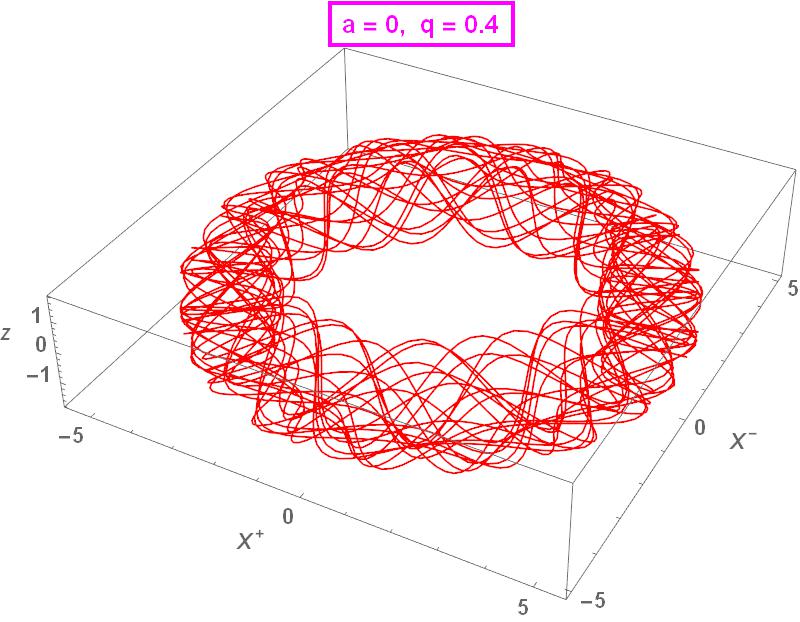}
\\
(i)\hskip58mm (ii)
\vspace{-3mm}
\caption{
\textit{\small Motion in a $3D$ Paul Trap for (i) $a=0.1$ (axial symmetry) (ii) $a=0$ (periodic profile).
The initial conditions $X^{+}(0)=0,\, \dot{X}^{+}(0)= 1,\, X^{-}(0)= 5,\, \dot{X}^{-}(0)= 0,\, z(0)=0,\, \dot{z}(0)=1$.}
\label{3Dtrajplot}
}
\end{figure}

Real traps are $3$-dimensional~: 
 ions are Paul-trapped by a time-dependent
 quadrupole potential, written, in appropriate units, as \cite{PaulNobel,Blumel},  
\beq
\Phi=\frac{1}{2}\big(a+2q\cos 2U\big)
\Big((X^{+})^2+(X^{-})^2-2z^2\Big),
\label{3DTrap pot}
\eeq
where $a$ and $q$ are parameters and we used again the notation $U=\omega t/2$.
(\ref{3DTrap pot}) is clearly an axi-symmetric anisotropic oscillator potential with time-dependent frequencies. 
The motion of an ion is described therefore by three uncoupled Mathieu equations,
\besub
\begin{align}
(X^{\pm})^{\prime\prime}+\,\big(a+2q\cos 2U\big)\,X^{\pm}&=\;0,
\label{Lxy}
\\
\quad z^{\prime\prime}\; -\; 2\big(a+2q\cos 2U\big)\,z&=\;0\,.
\label{Lz}
\end{align}
\label{3DTrap eq}
\esub
The interaction in the $X^{\pm}$ plane is attractive, while the one in the $z$ direction is repulsive and has a factor $2$. The oscillating term produces bounded motions in an appropriate range of parameters. For details the reader is referred to the  literature, e.g. \cite{Blumel}. Some bounded trajectories are shown in Figs.\ref{3Dtrajplot}.

The $3D$ Paul Trap can again be  lifted to Bargmann space  -- but one in $5D$. The recipe  is the same as before \cite{Bargmann}: the Bargmann metric is (\ref{Brinkmetric}) but now we have $3$ transverse components ; the $UU$ component is $-2\Phi(X^{\pm},z,U)$ given in (\ref{3DTrap pot}).
The quadratic form is traceless and therefore the metric still satisfies the vacuum Einstein equations
$R_{\mu\nu}=0$~: \emph{it is a gravitational wave in $5D$}.

\subsection{Penning Trap }\label{PennSec}

A similar however different ion trap was proposed by Dehmelt who called it the \emph{Penning Trap} \cite{DehmeltNobel,Brown,Blaum} (and who shared  for it the Nobel prize with Paul). It combines
an anisotropic but \emph{time-independent quadrupole potential} with a \emph{uniform (constant) magnetic field $\bB=B\,\hat{z}$
 directed along the $z$ axis},
\besub
\begin{align}
\Psi(Y^+,Y^-,z)&= - \big(\frac{\omega_z}{2}\big)^2\,\Big((Y^{+})^2+(Y^{-})^2-2z^2\Big), 
\label{PennPot}
\\
A_+&= -\half B {Y^-},\quad
A_-= \half B {Y^+},\quad
A_z= 0\,.
\label{PennVecpot}
\end{align}
\label{Pennpotvecpot}
\esub
The Lagrangian 
\beq
L=\frac{1}{2}\dot{\bY}^2+\frac{\omega_c}{2}(\dot{Y}^- Y^+-\dot{Y}^+ Y^-)+\frac{1}{4}\omega_z^2 (\bY^2-2z^2)\,,
\label{PNLag}
\eeq
where $\omega_c= B$ in our units is the cyclotron frequency \footnote{Once again, the dot means here $d/dt$ where $t$ is non-relativistic time.}, yields, for a particle of unit charge and mass, \besub
\begin{align}
& \ddot{Y}^{\pm} \mp \omega_c \dot{Y}^{\mp} - \dfrac{1}{2} \omega_z^2 \,{Y^{\pm}} = 0\,
\label{Penningxy} 
\\
& \ddot z + \omega_z^2 z = 0\,
\label{Penningz}
\end{align}
\label{Penningeq}
\esub
cf. eqns. \# (2.5)-(2.7) of ref. \cite{Kretzschmar}.

\begin{figure}[h]
\begin{center}
\includegraphics[scale=.27]{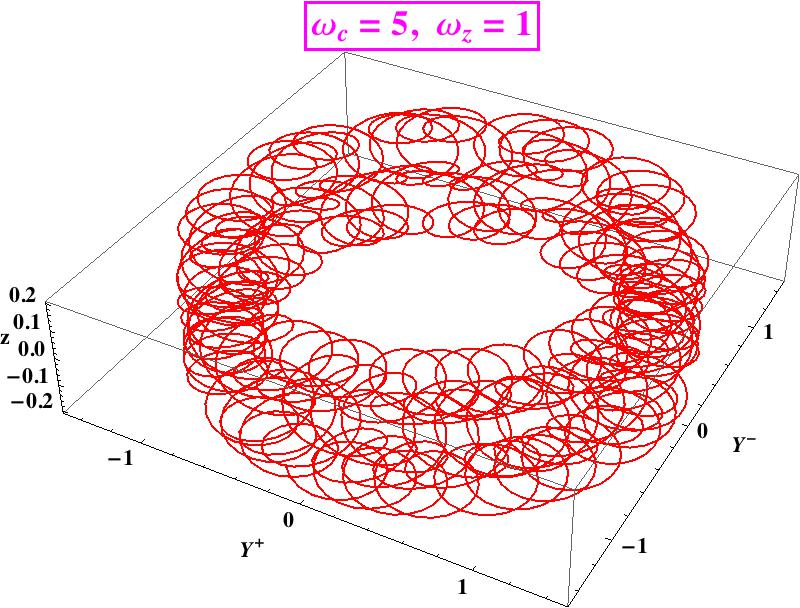}\qquad\quad
\includegraphics[scale=.2]{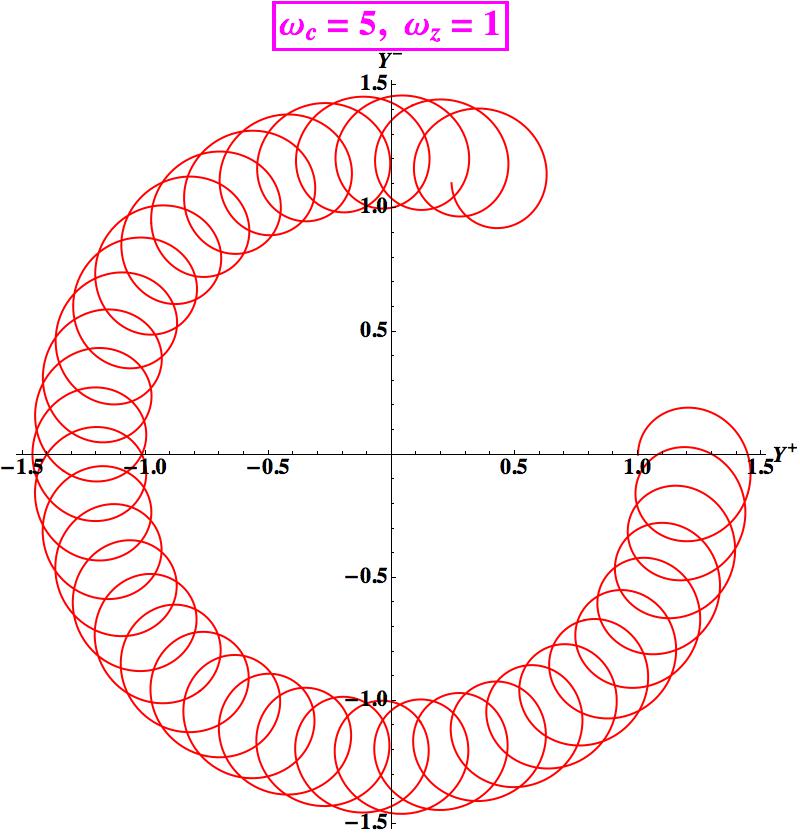}
\\ \hskip9mm
(i)\hskip74mm (ii)\\
\vskip-4mm
\caption{\textit{\small Trajectory of a charged particle in a Penning  Trap  (i) in $3D$ (ii) its projection on the $Y^{\pm}$ plane. The initial conditions are $Y^+(0)= 1.0,~\dot{Y}^+(0)=0.0,~Y^-(0)=0,~\dot{Y}^- (0)=1.0,~z(0)=0,~\dot z(0)=0.2$.}
\label{penningfig}}
\end{center}
\end{figure}

\begin{figure}[h]
\begin{center}
\includegraphics[scale=.28]{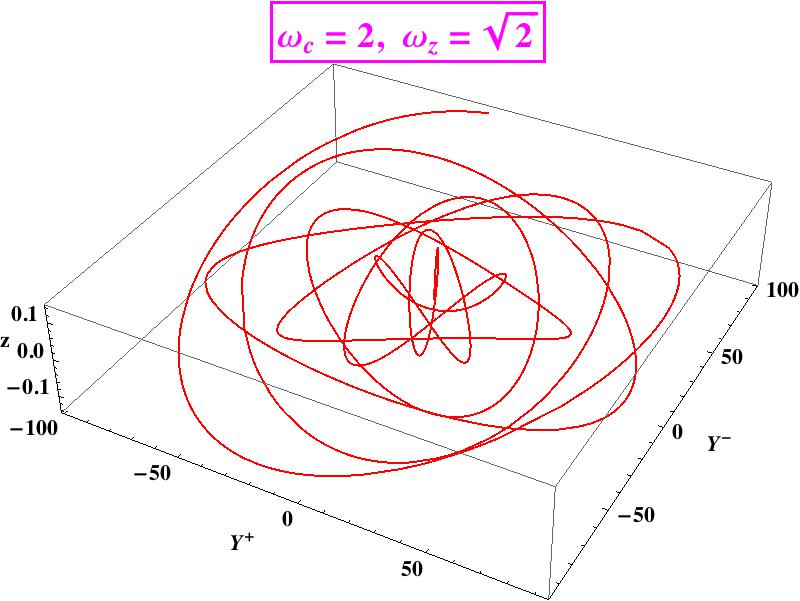}
 \qquad
\includegraphics[scale=.2]{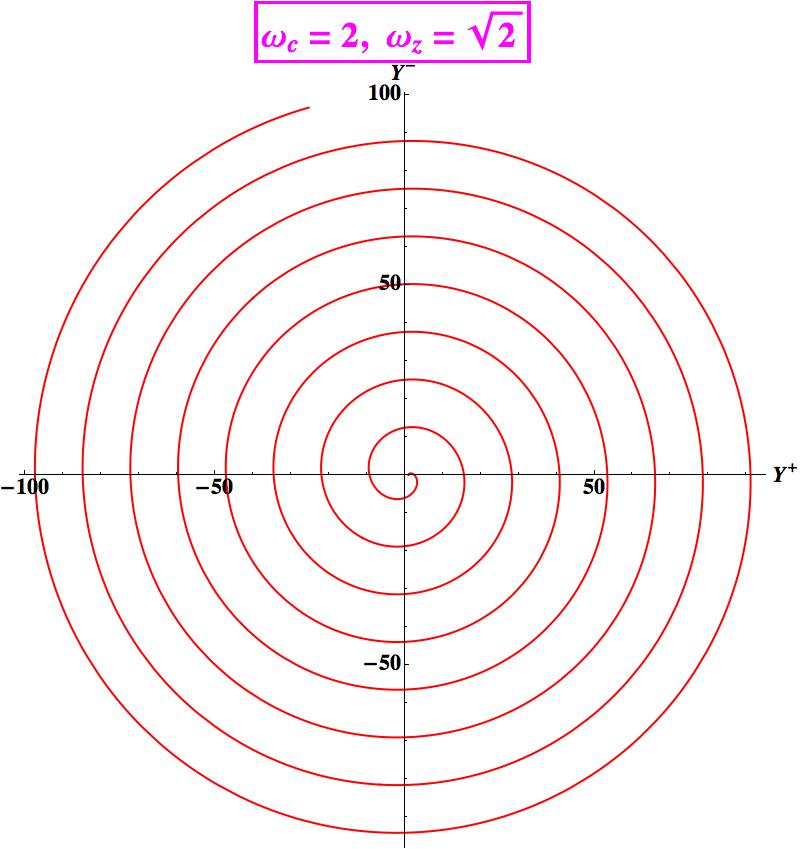}
\\ \vskip-2mm\hskip8mm
(i)\hskip63mm (ii)\\
\vskip-4mm
\caption{\textit{\small 
In the fine-tuned case  (\ref{specPenning}) and for
 initial conditions  $Y^+(0)= 1.0,~\dot{Y}^+(0)=0.0,~Y^-(0)=0,~\dot{Y}^- (0)=1.0,~z(0)=0,~\dot z(0)=0.2$, the $2d$ projection (ii) of the $3d$ trajectory (i) spirals outward with expanding radius.
The $z$ coordinate oscillates with frequency $\omega_z=\omega_c/\sqrt{2}$.}}
\label{Pspiral}
\end{center}
\end{figure}

Let us observe for further reference that the upper eqns, (\ref{Penningxy}),
are \emph{reminiscent of the circularly polarized periodic} (CPP) form (\ref{Yeqn}) [as suggested by our notations], while the $z$-equation is that of a decoupled harmonic oscillator. 
 In terms of the complex coordinate $\Upsilon=Y^{+}+iY^{-}$   eqn. (\ref{Penningxy}) is solved by \cite{Kretzschmar},
\beq
\Upsilon_{\pm}(t)=e^{-i\gamma_{\pm} t},
\qquad
\gamma_{\pm}=\half\big(\omega_c\pm\sqrt{\omega_c^2-2\omega_z^2}\big).
\label{pennsol}
\eeq
The constants $\gamma_+$ and $\gamma_-$ here are the modified cyclotron frequency and the magnetron frequency, respectively.
Periodic solutions require 
$\omega_c^2-2\omega_z^2>0$.
 Solutions are shown in Fig. \ref{penningfig} ; bound motions arise when $\omega_z^2>0$. 
In experimentally realistic cases $\omega_c >> \omega_z$ \cite{Brown,Kretzschmar}. 
However a special case arises
when the Penning trap has equal modified-cyclotron and  magnetron frequencies,
\beq
\gamma_+=\gamma_-=\half\omega_c,
\qquad\text{i.e.\ for}\qquad
\Delta =
\left(\frac{\omega_z}{\omega_c}\right)^2-\frac{1}{2}=0.
\label{specPenning}
\eeq
Then both motions in (\ref{pennsol}) coincide (are purely cyclotronic) whereas the new independent solution spirals outward  as shown in Fig.\ref{Pspiral}, reminiscent of the maximally anisotropic case $\Omega_-=0$ in Fig.9 of \cite{POLPER}, 
\beq
\Upsilon_{0}^{(per)}(t)=e^{-i\half\omega_{c} t},
\qquad
\Upsilon_{0}^{(esc)}(t) =t\,e^{-i\half\omega_{c} t}\,. 
\label{specPenningsol}
\eeq 
The toroidal region shrinks to a circle and we get also a new, escaping solution. 
The general solution of the  $3D$ system (\ref{Penningeq}), a combination of those in (\ref{pennsol}) completed with 
$ 
z=E\cos (\omega_z \,t) +F\cos (\omega_z \,t)
$, can be also obtained by chiral decomposition.  

For a discussion of the quantum aspects the reader  is referred, e.g. \cite{Brown} to for details. Here we just mention that the spectrum is \cite{Brown}, 
\beq
E_{(n_+, n_-, k)}=\hbar\bigg[
\gamma_+ \big(n_++\half\big) 
- \gamma_- \big(n_-+\half\big)
+\omega_z\big(k+\half\big)\bigg],
\qquad n_\pm, k= 0,1, \dots
\label{PNspectrum}
\eeq
as it can also be confirmed by the chiral method.
In the special case (\ref{specPenning}) the bound-state spectrum is that of the $z$ component alone, consistently with (\ref{specPenningsol}) and Fig.\ref{Pspiral}\,.

Now we turn to the GW aspect of 3D traps.
As said above, Paul Traps correspond to \emph{linearly polarized} periodic (LPP) waves; now we inquire if
\textit{the analogy can be extended by relating the Penning trap to CPP waves.}
We first recall how a non-relativistic particle in an external electromagnetic field can be described by a $5D$ Bargmann space  \cite{DHP2}. In terms of the coordinates $(\bY,z,t,s)$ we have,
\besub
\begin{align}
ds^2=
d\bY^2+dz^2+2dt\big(ds+A_idY^i\big)-2\Psi\, dt^2,
\end{align}
\label{Penmetric}
\esub
whose null geodesics project consistently with (\ref{Penningeq}).
Note that  the metric is \emph{not} Ricci-flat~: the potential (\ref{PennPot}) is harmonic, $\Delta\Psi=0$ and therefore
$ 
R_{UU}=-\half B^2 \neq0, 
$ 
cf. (\ref{ppRicci}). The  metric (\ref{Penmetric}) is thus not vacuum Einstein.

To get further insight, we now eliminate the vector potential in (\ref{Penmetric}) by  the rotational trick (\ref{rotframe}) [backward] extended to $3D$,
\beq
\barray{c}X^+\\ X^-\\ z\earray
=
\barray{rccc}\cos \omega t &&-\sin \omega t &0\\
\sin \omega t &&\cos \omega t &0
\\
0 &&0 &1\earray
\barray{c}Y^+\\ Y^-\\ z\earray
\label{3Drottrick}
\eeq
where $\omega$ is a constant.
The cross terms $dX^{\pm}dt$  cancel if  
$ \omega= {\omega_c}/{2}$ and we end up with \vspace{-3mm}
\besub
\begin{align}
ds^2 &=d\bX^2+ 2dtds-2\Phi\, dt^2,
\label{rotPennmetr}
\\[6pt]
\Phi &=
\dfrac{1}{8}({\omega_c^2}-2{\omega_z^2})
\Big[(X^+)^2 + (X^-)^2\Big]
+\dfrac{1}{2}\omega_z^2z^2\,,
\label{rotPennpot}
\end{align}
\label{rotPenn}
\esub
which is the Bargmann metric of an axially symmetric [attractive or repulsive, generally anisotropic] oscillator \footnote{For the special value (\ref{specPenning}) the oscillator is maximally anisotropic: $X$-motion is  free. Another extreme case would be $\omega_z=0$ when the $z$-motion is free. When $\omega_c^2=6\,\omega_z^2,$ 
the $X$-oscillator (\ref{rotPennpot}) is isotropic.
}. 
Therefore, \emph{despite the similarity between the upper two Penning eqns (\ref{Penningxy}) and the CPP equations
(\ref{Yeqn}), the Bargmann lift of a Penning trap is not a CPP GW~: it is not Ricci-flat (as confirmed again by $\Delta \Phi = {\omega_c^2}/2\equiv B^2/2$) and is not brought to the CPP form by the rotational trick.}

\subsection{Modified Penning trap}\label{PMtrapSec}
 
Below we propose instead a \emph{modified Penning trap}, closer to  CPP GWs.
We first note a subtle however important difference between the two systems~: in (\ref{Penningxy})
the $Y^{\pm}$ terms  have identical frequencies $\omega_z$, whereas in the CPP case (\ref{Yeqn}) the frequencies are different, $\Omega_+^2\neq\Omega_-^2$, except when $A_0=0$ --- i.e., when there is no wave. Therefore we propose to generalize the scalar Penning potential (\ref{PennPot}) while keeping the same vector potential (\ref{PennVecpot}),
\besub
\begin{align}
&\medbox{
\Psi \to \widetilde{\Psi} = -(\dfrac{\omega_z}2)^2 \left((1+ \dfrac{A_0}{2})(Y^+)^2+ (1-\dfrac {A_0}{2}) (Y^-)^2 - 2 z^2\right)
} 
\label{modPennPot}
\\[4pt]
&A_{\pm} = \mp\half\,\omega_c Y^{\mp},\;\; A_z=A_t=0\,,
\label{vecPot}
\end{align}
\label{PMtrap}
\esub
where $A_0$ is a perturbation parameter.
The new term clearly breaks the axial symmetry whenever $A_0\neq0$.
Spelling out for completeness, the Lagrangian 
\beq
L\!=\!\frac{1}{2}(\dot{\bY}^2+\dot{z}^2)+\frac{\omega_c}{2}(\dot{Y}^- Y^+-\dot{Y}^+ Y^-)
+\frac{\omega_z^2}{4}\left(\Big(1+\frac{A_0}{2}\Big)(Y^+)^2+\Big(1-\frac{A_0}{2}\Big)(Y^-)^2
-2z^2\right),\qquad
\label{PMLag}
\eeq
cf. (\ref{PNLag})
yields the  equations of motions, 
\beq
\label{PMeqn}
\ddot{Y}^\pm \mp \omega_c\dot{Y}^\mp - \frac{\omega_z^2}{2}\Big(1\pm \frac{A_0}{2}\Big)Y^\pm=0, 
\qquad
\ddot{z}+\omega_z^2 z=0.
\eeq
cf. (\ref{Penningeq}). Lifting to $5D$ Bargmann space $(Y^+,Y^-,z,t,s)$, our modification amounts to considering
\beq
ds^2 = (dY^+)^2+(dY^-)^2+(dz)^2+2dt(ds+A_i dY^i) -2\widetilde{\Psi} dt^2\,
\label{modPennBarg}
\eeq
where the vector potential is  still 
 (\ref{PennVecpot}).
Then applying once again the $3D$ rotational trick (\ref{3Drottrick}) allows  us to conclude along the same lines as above that choosing $\omega =B/2\equiv \omega_c/2$  and putting 
$U=\omega_c t/2, \, V=2s/\omega_c$,
we get
\besub
\begin{align}
ds^2 &=  (dX^+)^2+(dX^-)^2 +(dz)^2 + 2dUdV -2\widetilde{\Phi}dU^2\,,
\\[8pt]
& \widetilde{\Phi} = \left(
\dfrac{1}{2}-\big(\dfrac{\omega_z}{\omega_c}\big)^2 \right)
\Big[(X^+)^2+(X^-)^2\Big]+2 (\dfrac{\omega_z}{\omega_c})^2z^2
\nonumber 
\\[4pt]
&\;\;\quad - \big(\dfrac{\omega_z}{\omega_c}\big)^2\,\dfrac {A_0}{2} 
\bigg[\cos 2U \left( (X^+)^2 - (X^-)^2 \right) + 2 \sin 2U(X^+X^-)\bigg], 
\end{align}
\esub
which is a rather complicated mixture of a time-dependent oscillator with a periodic correction term. However when 
\beq
\Delta =
\left(\frac{\omega_z}{\omega_c}\right)^2-\frac{1}{2}=0\,,
\label{specomegacz}
\eeq
cf. (\ref{specPenning}), the isotropic part is turned off, leaving us  with a CPP GW embedded into $5D$
Bargmann space, 
\beq\bigbox{
\widetilde{\Phi}_{spec} = 
- \half \widetilde{K}_{ij}X^i X^j =
- \dfrac{A_0}{4}\bigg[\cos 2U \Big((X^+)^2 - (X^-)^2\Big) + 2\sin 2U \Big(X^+ X^-\Big)\bigg]+z^2\,,
}
\label{specPennPot}
\eeq
which identifies the constant $A_0$ as the amplitude of the CPP GW in 5D, the Bargmann space of the modified Penning trap. For $A_0=0$ we recover the maximally anisotropic Penning case 
$\Phi=({\omega_c^2}/{4})z^2$, cf. (\ref{rotPennpot}).
In the special case (\ref{specomegacz}), 
the chiral decomposition of the system (\ref{PMLag})-(\ref{PMeqn}) is found as, 
\besub
\begin{align}
H&=H_+- H_-+H_z 
\label{pmchirH}
\\[4pt]
&=\frac{1}{2}\left[\frac{\omega_c^2 A_0}{8}\Big((1+A_0/2)Z_+^1 Z_+^1 + Z_+^2 Z_+^2 \Big)
-\frac{\omega_c^2 A_0}{8}\Big((1-A_0/2)Z_-^2 Z_-^2 + Z_-^1 Z_-^1\Big)+p_z^2+\frac{\omega_c^2}{2}z^2 \right],
\nn
\\[12pt]
\sigma&=\sigma_+- \sigma_-+\sigma_z=-\frac{\omega_c A_0}{4}\left[dZ_+^1\wedge dZ_+^2-dZ_-^1\wedge dZ_-^2 \right]+dp_z\wedge dz. 
\label{pmchirs} 
\end{align}
\label{PMchiraldec}
\esub
cf. (\ref{polperdd})-(\ref{polpersH}).
The resulting uncoupled equations,
\beq
\ddot{Z}_{\pm}^{1,2}+\frac{\omega_c^2}{4}(1{\pm}A_0/2) Z_{\pm}^{1,2}=0, 
\qquad
\ddot{z}+\frac{\omega_c^2}{2} z=0\,,
\eeq
 are solved at once ; in $Y$-coordinates, we get,
\besub
\begin{align}
Y^+&= A \cos\left(\frac{\omega_c}{2}\sqrt{1+\frac{A_0}{2}}\,t\right)+B \sin\left(\frac{\omega_c}{2}\sqrt{1+\frac{A_0}{2}}\,t\right)\nn
\\[6pt]
&+C\cos\left(\frac{\omega_c}{2}\sqrt{1-\frac{A_0}{2}}\,t\right)+D \sin\left(\frac{\omega_c}{2}\sqrt{1-\frac{A_0}{2}}t\right)\quad \quad 
\\[12pt]
Y^-&= \sqrt{1+\frac{A_0}{2}} \left[ B \cos\left(\frac{\omega_c}{2}\sqrt{1+\frac{A_0}{2}}\,t\right)-A \sin\left(\frac{\omega_c}{2}\sqrt{1+\frac{A_0}{2}}\,t\right) \right] 
\nonumber
\\[6pt]
&+ \frac{1}{\sqrt{1-\displaystyle\frac{A_0}{2}}}\left[D\cos\left(\frac{\omega_c}{2}\sqrt{1-\frac{A_0}{2}}t\right)-C \sin\left(\frac{\omega_c}{2}\sqrt{1-\frac{A_0}{2}}\,t\right) \right]\,, 
\end{align}
\label{PMsol}
\esub
The quantum spectrum can be obtained using creation/annihilation operators, 
\beq\bigbox{
E_{n_+, n_-, k}=\hbar\bigg[\sqrt{1+A_0/2}\, \Big(n_++\half\Big) 
- \sqrt{1-A_0/2}\, \Big(n_-+\half\Big)
+ \sqrt{2}\,k+\frac{1}{\sqrt{2}}\bigg]
\,}
\label{PMspectrum}
\eeq
where $n_\pm= 0,\ 1\dots$ are the eigenvalues of the appropriate number operators, and $k=0, 1$ is that of the $z$-oscillator \footnote{$\omega_c=2$ in our units.}. For a weak wave, $A_0<<1$, we have,
\beq
E_{n_+, n_-, k} \approx \hbar\Big[(n_+-n_-) +\frac{A_0}{4}
\big(1+ (n_++n_-)\big)+ \sqrt{2}\,k+\frac{1}{\sqrt{2}}\Big]\,.
\label{weakPMspectrum}
\eeq

\section{Lagrange points in Celestial Mechanics}\label{3bodysec}

In \cite{BBLag,BBTroj}  Bia\l ynicki-Birula et al discuss the  stability of Lagrange points 
 in the Newtonian 3-body problem using a linearized Hamiltonian \cite{Danby}. In the co-rotating $x-y$ plane defined by the two main orbiting bodies the Hamiltonian takes the form 
\beq 
\label{eq:BB_Hamiltonian}
H_{osc} = \frac{p_x^2 + p_y^2}{2} + \frac{a\, \omega^2 x^2 + b\, \omega^2 y^2}{2} - \omega (x p_y - y p_x) \, , 
\eeq 
where the values of $\omega$ and the dimensionless $a$ and $b$ depend on the parameters of the original problem. The authors discuss in particular islands of stability in the space of parameters. The equations of motion arising from \eqref{eq:BB_Hamiltonian} are\footnote{The ``one-sided'' ``Hill'' case studied in \cite{ZGH}
corresponds to $a=-2$ and $b=1$ and was found unstable.}, 
\beq 
\ddot{x} - 2 \omega \dot{y} = \omega^2 (1-a) x \, , \qquad
\ddot{y} + 2 \omega \dot{x} = \omega^2 (1-b) y \, .
 \label{eq:BB_eoms}
\eeq 

The values of $a$ and $b$ can be found by comparing with the results in textbooks such as \cite{Danby}. In this reference units are chosen so that distances, time, and masses are expressed by dimensionless quantities. Distances are measured from the center of mass of the two main rotating bodies, and rescaled by their relative distance. 
 In the co-rotating frame the two rotating bodies lie on the $x$ axis, and the unit of time is chosen so that the angular velocity of rotation of the co-rotating frame is $\omega = 1$. Our ``big masses'' are labeled so that $M_1 \le M_2$\,, which implies that
\beq
0< \mu \equiv \frac{M_1}{M_1 + M_2}
\leq 1/2\,.
\label{murange}
\eeq

\begin{figure}
\begin{center}
\includegraphics[scale=.51]{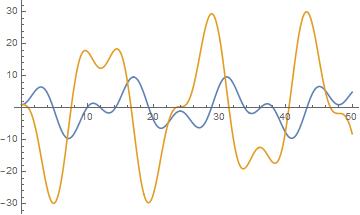}
\qquad
\includegraphics[scale=.53]{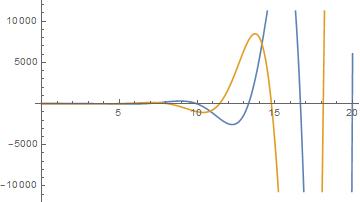}
\vspace{-3mm}
\caption{
\textit{\small Motions around a Lagrange point are bounded for $\mu= 0.02<1/27$ and unbounded for $\mu= 0.4>1/27$.
}
\label{Lagplot}
}
\end{center}
\end{figure}

Here we are  interested in the two Lagrangian points,
traditionally denoted by $L_4$ and $L_5$.
The displacements  $\xi$, $\eta$  in the $x - y$ plane  satisfy the coupled  equations : 
\besub
\begin{align} 
\ddot{\xi} - 2 \dot{\eta} = \frac{3}{4} \xi + \frac{3\sqrt{3}}{4} (1 - 2 \mu) \eta \, ,
 \\ 
\ddot{\eta} + 2 \dot{\xi} = \frac{9}{4} \eta + \frac{3\sqrt{3}}{4} (1 - 2 \mu) \xi \, , 
\end{align}
\esub 
from which we can read off a scalar potential 
$$ 
V (\xi, \eta) = - \frac{3}{8} \xi^2 - \frac{3\sqrt{3}}{4} (1-2\mu) \xi \eta - \frac{9}{8} \eta^2 \,, 
$$ 
whose Hessian 
 $\p_i\p_j{V}$ has eigenvalues 
\beq 
\lambda_{1,2} = \frac{3}{2}\left(\, - 1 \mp \sqrt{1- 3 \mu + 3 \mu^2}\, \right)= 
\frac{3}{2}\left(\, - 1 \mp \sqrt{3(\mu-\half)^2+\smallover{1}/{4}}\, \right).  
\label{HessianEigen}
\eeq
The potential can be diagonalised by a rotation, 
$ 
{\tiny \left( \begin{array}{c} \hat{\xi} \\ \hat{y} \end{array} \right)} = R {\tiny \left( \begin{array}{c} \xi \\ \eta \end{array} \right)} \,  
$  
which brings the equations of motion to the form,\vspace{-5mm}
\besub
\begin{align} 
\ddot{\hat{\xi}} - 2 \dot{\hat{\eta}} = - \lambda_1 \hat{\xi} \, , 
\\ 
\ddot{\hat{\eta}} + 2 \dot{\hat{\xi}} = - \lambda_2 \hat{\eta} \, . 
\end{align}
\label{Eigeneqmot}
\esub 
Then by  comparison with (\ref{eq:BB_eoms}) we get 
$  
\omega = 1, \, 
\lambda_1 = a - 1 \, , \, 
\lambda_2 = b - 1 \, . 
$ 
Eqn. (\ref{HessianEigen})  implies
$  
 a + b = - 1 \,  
$ 
cf. \cite{BBLag},
which allows us to infer that
\beq
a = - \frac{1}{2} - \frac{3}{2} \sqrt{1 - 3 \mu +3 \mu^2} \, ,
\qquad
b = - \frac{1}{2} + \frac{3}{2} \sqrt{1 - 3 \mu +3 \mu^2} \, ,
\label{abmu}
\eeq 
A standard argument for stability goes along these lines: deriving the equations (\ref{Eigeneqmot}) twice and using (\ref{HessianEigen}) allows us to eliminate either $\hat\xi$ or $\hat\eta$ (say $\hat\eta$). Using that 
$\lambda_1+\lambda_2=-3$ and 
$\lambda_1\lambda_2=(27/4)\mu(1-\mu)$  by 
(\ref{HessianEigen}), we get the fourth-order eqn
$$
\hat{\xi}^{(iv)}+\ddot{\hat{\xi}}+\smallover{27}/{4}\mu(1-\mu)\,\hat{\xi}=0.
$$
The simple exponential $\hat\xi(\tau)=e^{\alpha \tau}$ is a solution if
$
\alpha^4 +\alpha^2 +\frac{27}{4}\mu(1-\mu)=0.
$
When 
\beq
\mu(1-\mu)< \frac{1}{27}\, 
\label{BBstab}
\eeq
holds, then  
$\alpha^2$ is real and is in fact negative, yielding a purely imaginary $\alpha$~: the solution is periodic  \footnote{
For the Sun-Jupiter system 
$\mu \approx 10^{-3}$ yielding $\mu(1-\mu) << 1/27 \approx 0.037037$,  consistently with the observed stability of the ``Greek/Trojan'' minor planets (asteroids). 
For the Earth-Moon system $\mu \approx 0.012\,\Rightarrow \mu(1-\mu) \approx 0.0118 < 1/27$. The discovery of the first Earth-Trojan, 2010 TK7, was announced by NASA in 2011.
 For the Sun - Earth system $\mu \approx \mu(1-\mu) \approx 3.10^{-6}$ -- stable.
This has its importance for, say, LISA, with GW detectors planned to be sent
to the Lagrange points.}. The condition above appears in \cite{BBLag}, and is  confirmed by numerical integration, cf. Fig.\ref{Lagplot}. 
 
The reader might have noticed that the equations (\ref{Eigeneqmot}) are mapped into our previous equations (\ref{Yeqn}) by replacing the $-\lambda_i$ with $\Omega_{\pm}^2$. For the latter set of equations we found stability (trigonometric functions) provided that the frequency- squares are both positive, making  seemingly unnecessary the stability condition (\ref{BBstab}). In sec.\ref{PerGravWsec} however we have worked with the specific condition $\frac{1}{2}\left(\Omega_{+}^2 + \Omega_{-}^2 \right) = 1$, which does not apply here, so that the solutions (\ref{polperLsolns}) can not be used directly. Setting $ c=\sqrt{1-3\mu(1-\mu)}$
 and 
$\Delta=9c^2-8=1-27\mu(1-\mu)$, the chiral decomposition (\ref{polperdd}) has coefficients 
\begin{eqnarray}
\alpha_\pm = \frac{1}{4}(4-3c\mp \sqrt{\Delta})
, \quad \beta_\pm = \frac{1}{4}(4+3c\mp \sqrt{\Delta}),
\label{lagms}
\end{eqnarray}
which are real when 
$
\Delta\geq0
$, i.e., when
$
\mu(1-\mu) \leq \frac{1}{27}\,.
$
Then the Hamiltonian $H$ and the symplectic $2-$form $\sigma$ are decomposed as
\besub
\begin{align}
&H=
\frac{\sqrt{\Delta}}{16}\left[
\big(\sqrt{\Delta}-4+3c\big)X^2_+X^2_+
+
\big(\sqrt{\Delta}-4-3c\big)X^1_+X^1_+\right.
\\
&\qquad\left.
+
\big(\sqrt{\Delta}+4-3c\big)X^2_-X^2_-
+
\big(\sqrt{\Delta}+4+3c\big)X^1_-X^1_-
 \right],
\label{(2a)} 
\\[6pt]
&\sigma=
\frac{\sqrt{\Delta}}{2}
\left[dX^1_+\wedge dX^2_+ - dX^1_-\wedge dX^2_- \right] \, , 
\label{(2b)}
\end{align}
\label{(2ab)}
\esub
whose regularity requires 
$
\Delta\neq0$. In conclusion, the condition (\ref{BBstab}) should hold. Then the corresponding solutions, 
\besub
\begin{align}
&\hat{\xi}= A\cos(\sqrt{\alpha_+\beta_+}\,t)+ B\sin(\sqrt{\alpha_+\beta_+}\,t) \nn
\\
&\hskip6mm
+\, C\cos(\sqrt{\alpha_-\beta_-}\,t)+D\sin(\sqrt{\alpha_-\beta_-}\,t), 
\label{(4a)}
\\[8pt]
&\hat{\eta}= \sqrt{\frac{\beta_+}{\alpha_+}}\left(-A\sin(\sqrt{\alpha_+\beta_+}\,t)+ B\cos(\sqrt{\alpha_+\beta_+}\,t) \right)\nn
\\
&\hskip6mm
+\, \sqrt{\frac{\beta_-}{\alpha_-}}\left (-C\sin(\sqrt{\alpha_-\beta_-}\,t)+ D\cos(\sqrt{\alpha_-\beta_-}\,t) \right) 
\label{(4b)} 
\end{align}
\esub 
are manifestly bounded, because  $\Delta < 1$ implies
$
\alpha_\pm\beta_\pm=\frac{1}{2} \left(1\mp \sqrt{\Delta} \right)>0.
$

\section{Conclusion}

The striking similarities of the seemingly far remote topics discussed in this paper have, from the mathematical point of view, a simple explanation: in all cases, the problem boils down to  study an \emph{anisotropic oscillator } \cite{Alvarez,ZGH,ZHAGK}.
In the Paul [alias linearly polarized GW] case, 
the solution is expressed in terms of Mathieu functions \cite{PaulNobel,Harte,POLPER}; in the circularly polarized periodic 
GW case, they involve trigonometric/hyperbolic functions.

Previous investigations of the memory effect 
 focused on sudden bursts of \emph{sandwich waves} which vanish outside a short ``wave zone". It was advocated \cite{velmem} that their observation would (theoretically) be possible due to the \emph{velocity memory effect}~: in the flat ``afterzone'' the particles move indeed with constant velocity, as required by  \dots Newton's 1st law  \cite{ShortLong,ImpMemory,POLPER}.

In this paper we study instead \emph{periodic waves} sought for in inflationary models \cite{Bmode,LIGOPer}. Such waves have no ``before and after-zone", and their observation would require a different technique.
Here we argue that, by analogy with ion trapping \cite{PaulNobel,DehmeltNobel}, one might study bound motions. 

The Eisenhart lift of the 3D Paul trap is a linearly polarized periodic gravitational wave in 5D. For 3D Penning traps we find that, despite strong similarity  with the equations which govern circularly polarized gravitational waves, their  Eisenhart lift   is \emph{not} a CPP wave. However a slight modification  (see sec. \ref{PMtrapSec}) allows for  anisotropy and for a special value (\ref{specomegacz}) of the frequencies we do get circularly polarized gravitational waves in 5D. 
Such perturbations were actually considered before as due to imperfections, see eqn. \# (2.71) of ref. \cite{Brown}. 

It is remarkable  that molecular physicist who  worked on ion traps decades ago, were, like Moli\`ere's \textit{Monsieur Jourdain}, studying gravitational waves.

While our investigations here are classical, 
 the chiral decomposition,  (\ref{polperH}), makes it easy to  study the quantum problem. 
 Observing  in particular the bound-state {spectrum} (\ref{PMspectrum})  could, theoretically, lead to the detection of such a wave.
We mention that ion traps have also been studied recently in connection with (space)time crystals \cite{STcrystals}. 
\goodbreak

It is worth to emphasize that these kinds of analogies extend very generally, even for non-periodic waves: the geodesic deviation equation in a vacuum background always looks like $\ddot{\xi}= R \xi$ in a parallel-propagated frame, where $R$ is an appropriate matrix. But this is an anisotropic oscillator equation and such equations are ubiquitous in physics. Moreover, time dependence in $R$ can give
parametric resonance and similar things, as is well-known in other contexts. Geodesic motion in curved spacetimes is therefore generically linked to time-dependent anisotropic oscillators. Plane wave spacetimes are special here because i) it's not awkward to let $R$ have any time dependence you like, and ii) the geodesic deviation equation is exact even for finite separations.

\begin{acknowledgments}
This paper is an \emph{hommage} in memory of our late friend and collaborator Christian Duval (1947-2018), who deceased just after its publication. 
PH is grateful to Piotr Kosinski and Csaba S\"uk\"osd for correspondence and to Matt Kalinski for pointing out an error (now corrected) in the published version of this paper, see our Erratum \cite{Erratum}. The last paragraph of our Conclusion has been copied \textit{ad verbatim} from the report of our anonymous referee to whom we express our indebtedness.
 ME and  PH want to thank the \emph{Institute of Modern Physics} of the Chinese Academy of Sciences in Lanzhou (China) for hospitality. This work was supported by the Chinese Academy of Sciences President's International Fellowship Initiative (No. 2017PM0045), and by the National Natural Science Foundation of China (Grant No. 11575254). MC acknowledges CNPq support from project (303923/2015-6), and a Pesquisador Mineiro project n.~PPM-00630-17.
\end{acknowledgments}

\goodbreak


\end{document}